\documentclass{JAC2003}
\usepackage{graphicx}
\usepackage{booktabs}

\begin{document}
\title{POST-LINAC COLLIMATION SYSTEM FOR THE EUROPEAN XFEL}

\author{V. Balandin\thanks{vladimir.balandin@desy.de}, R. Brinkmann, W. Decking, N. Golubeva \\
DESY, Hamburg, Germany}

\maketitle

\begin{abstract}
In this article we give an overview of the the post-linac collimation
system for the European X-Ray Free-Electron Laser (XFEL) Facility~\cite{XFEL}
with main emphasis on lattice and optics design. 
\end{abstract}

\section{INTRODUCTION}

The post-linac collimation system should simultaneously fulfill several 
different functions. In first place, during routine operations, it should 
remove with high efficiency off-momentum and large amplitude halo particles, 
which could be lost inside undulator modules and become the source of 
radiation-induced demagnetization of undulator permanent magnets. 

The system also must protect the undulator modules and other downstream 
equipment against miss-steered and off-energy beams in the case of machine
failure without being destroyed in the process. Because the collimation 
system is designed as passive, the collimators must be able to withstand 
a direct impact of such number of bunches which can be delivered to their 
locations until a failure will be detected and the beam production in 
the RF gun is switched off.

From the beam dynamics point of view, the collimation section, as a part of 
the beam transport line from linac to undulator, must meet a very tight set 
of performance specifications. It should be able to accept bunches with 
different energies (up to $\pm 1.5\%$ from nominal energy) and transport 
them without any noticeable deterioration not only of transverse, but also 
longitudinal beam parameters, i.e. it must be sufficiently achromatic and 
sufficiently isochronous. This will not only reduce jitter of transverse beam 
parameters and time of flight jitter due to an energy jitter, but also will 
allow to fine-tune the FEL wavelength by changing the electron beam energy 
without adjusting magnet strengths (an energy change of $\pm 1\%$ corresponds 
to $\pm 2\%$ change in the FEL wavelength) and, even more, will make possible 
to scan the FEL wavelength within a bunch train by appropriate programming of 
the low level RF system.  

Some of above requirements are not in good agreement with one another and,
as often, the basic problem is to find a balance among all competing factors 
so as to have at the end a system which still satisfactory fulfills design goals. 
For example, relatively large betatron functions, which are needed at the
collimator locations to guarantee their survival during occasional beam impacts,
lead, as a rule, to unacceptable chromatic aberrations and, therefore, 
chromaticity correcting sextupoles are essential in preventing the dependence
of linear optical parameters on the energy deviation. Chromatic-aberration 
correction with sextupoles, in the next turn, requires a beamline with dispersion, 
which makes separate regulation of transverse and energy collimation depths 
difficult and thus reduces flexibility of a system.

In this article we give an overview of the optics solution which fulfills 
all listed above requirements, and more details can be found in~\cite{ColXFEL}.

\section{LAYOUT AND FUNCTIONALITY}

\begin{figure}[htb]
    \centering
    \includegraphics*[width=75mm]{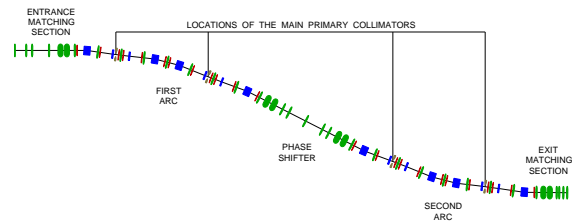}
    \caption{Overall layout of the XFEL post-linac collimation section.
    Blue, green and red colors mark dipole, quadrupole and sextupole magnets,
    respectively.}
    \label{figOverview}
\end{figure}

The part of the beam transport from linac to undulator, which we call the 
{\bf post-linac collimation section} and which is shown in Fig.~\ref{figOverview}, 
consists of two arcs separated by a straight section (phase shifter) and includes 
matching modules at both ends to adapt the optic to the desired upstream and 
downstream beam behavior. The collimation section bends the beam in the vertical 
plane and its length measured in the projection on the linac axis is about $215.3\,m$. 
The outgoing beam axis points slightly downward with an angle of about $0.021^{\circ}$
and the vertical offset of the center of the last quadrupole of the second arc from 
the linac axis is $2.4\,m$. The arcs are almost identical except that in the second 
arc the polarity of dipole and sextupole magnets is reversed and dipole bending 
angles are slightly smaller in absolute values in order to produce the net downward 
beam deflection. Each arc consists of four $90^{\circ}$ cells, constitutes a
second-order achromat and is first-order isochronous.

Three different types of collimators are foreseen in the XFEL post-linac collimation 
system: Main primary collimators, supplementing primary collimators and secondary 
collimators (absorbers). The principal purpose of the main primary collimators is 
to intercept trajectories of all incoming particles which would otherwise appear 
outside of the downstream dynamic aperture. The main collimators will also shade 
a part of the beam pipe in the collimation section (but not all) from uncontrolled 
beam impacts. Supplementing primary collimators assist to accomplish this work or, 
at least, reduce the probability of such events. Simultaneously, the supplementing 
collimators will not affect the transverse and energy collimation depths set by 
the main collimators. To improve the overall cleaning efficiency at the exit 
and better localize losses inside of the collimation section several absorbers 
placed in the shadow of the primary collimators will be used. 

The first arc will collimate transverse positions of incoming particles and the second, 
after a shift of vertical and horizontal betatron phases by odd multiples of $90^{\circ}$,
their transverse momenta. The energy and vertical plane collimation will be done 
simultaneously, and therefore the ratio of dispersion to vertical betatron function at 
collimator locations has to be properly adjusted in order to achieve the required 
transverse and energy collimation depths. Because, according to the optics design, 
dispersion can not be varied during machine operations, the rough preliminary adjustment 
was made already during design stage by appropriate selection of the arc parameters, 
and the operational flexibility will be provided by usage of collimators with exchangeable 
apertures and by tuning betatron functions at the collimator locations.

\section{OPTICS AND BEAM DYNAMICS}

Large beam spot size requested at the collimator locations and, in the same time, 
the possibility to transport bunches with different energies (up to $\pm 1.5\%$ 
from nominal energy) while preserving with good accuracy energy independent 
input and output matching conditions, make the control of chromatic effects 
one of the main issues in the design of the optics in the collimation section. 
Without correction the chromatic aberrations are unacceptable and, therefore, 
introduction of chromaticity correcting sextupoles becomes essential in improving 
overall system performance. There are different approaches to the problem of 
compensation of chromatic effects, and the solution, which we found to be most 
adequate to the design requirements, is as follows: We compensate the arc chromatic 
effects by tuning arcs to become second-order achromats. Reduction of chromatic 
aberrations in the system straight sections is done for the particular betatron 
functions transported through these parts and without involving sextupole fields, 
simply by an accurate drift-quadrupole optics design (a straight drift-quadrupole 
system can not be made an achromat, but it can be made a second-order apochromat
with respect to certain incoming beam ellipses, i.e. it can transport these beam 
ellipses without introducing first-order chromatic distortions).

\subsection{Adjustment of Linear Isochronicity}

\begin{figure}[htb]
    \centering
    \includegraphics*[width=75mm]{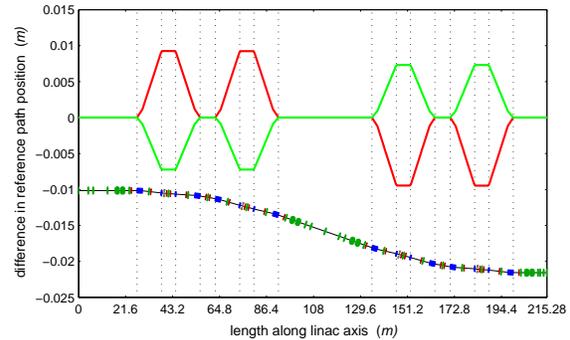}
    \caption{Changes in the vertical position of isochronous
    beam line with $r_{56}=0$, which are required to bring 
    linear momentum compaction of the collimation section to
    $r_{56}=-1\,mm$ (red curve) and to $r_{56}=+1\,mm$ (green curve).}
    \label{figDiffRefPath}
\end{figure}

In the first design, which is described in the TDR~\cite{XFEL}, the linear momentum 
compaction of the whole collimation system ($r_{56}\,$ matrix coefficient) was 
approximately equal to $-0.8\,mm$, which at that time was considered as acceptable
value. In later studies of the microbunching instability it was found that
even such a small value can not be neglected in the calculation of the gain of 
this instability and that, in order to reduce this gain, it is desirable to have 
$\,r_{56}\,$ of the collimation section equal to zero or, even better, to bring
it to a positive value of about $0.2\,mm$~\cite{Dohlus}. Because some other reasons 
for the choice of the linear momentum compaction could appear and the exact value 
of $r_{56}$ is not clear yet, we made system modifications in such a way, that 
though $\,r_{56}\,$ could not be varied dynamically during machine operations, 
the linear momentum compaction can be adjusted within about $\,\pm 1\, mm\,$ 
limits by system realignment while keeping space positions not only of the system 
end point and the system straight sections but also of the arc centers unchanged, 
as can be seen in Fig.~\ref{figDiffRefPath}. The design which we describe in this 
paper ({\bf baseline design}) is the first-order isochronous beam line with $r_{56}=0$.

\subsection{Linear Lattice Functions}

\begin{figure}[htb]
    \centering
    \includegraphics*[width=75mm]{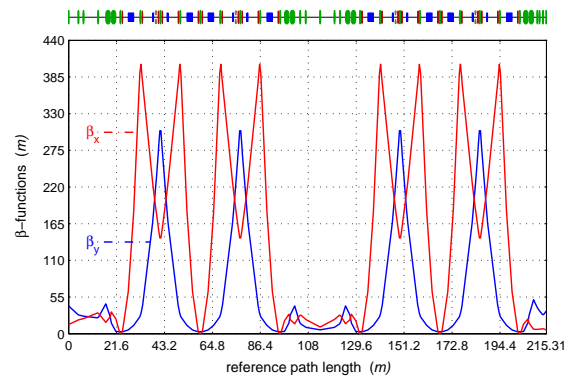}
    \caption{Betatron functions along the XFEL post-linac collimation section.
    Standard collimation optics.}
    \label{figBeta}
\end{figure}

\begin{figure}[htb]
    \centering
    \includegraphics*[width=75mm]{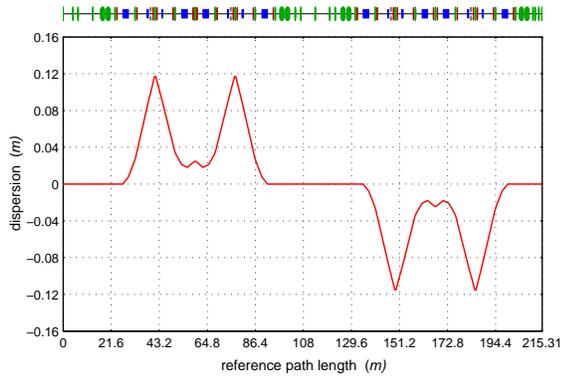}
    \caption{Dispersion function along the XFEL post-linac collimation section.}
    \label{figDisp}
\end{figure}

According to the system design the arcs are tuned to become second-order achromats 
and this, together with the fixed system geometry and the requirement of the 
first-order isochronicity, completely determines the setting of the arc magnets 
and also the behavior of linear dispersion function. Thus the modifications of 
the betatron functions along the collimation section can be provided only by 
retuning of quadrupoles in the matching sections and in the phase shifter. This is 
not a limitation in our case, because the design of these sections is done in such 
a way that with appropriate adjustment of their quadrupoles we are not only able 
to provide betatron functions of about $200\,m$ at the points where the collimators 
are located ({\bf standard collimation optics}), but also able to translate smoothly 
the standard collimation optics into an optics with regular FODO-like transport 
through the entire collimation section.

This flexibility is an important property of the designed system and will be 
extensively used during machine commissioning and/or during measurements of beam 
parameters. For example, commissioning starts with optics set to provide regular 
FODO-like transport and with sextupoles switched off and then, with experience gained, 
this optics can be translated step by step  into the standard collimation optics.

Betatron functions corresponding to the standard collimation optics can be seen in
Fig.~\ref{figBeta}. Fig.~\ref{figDisp} shows the linear dispersion, which is 
independent on setting of quadrupoles in the matching sections and in the 
phase shifter.

\section{SUMMARY}

\begin{figure}[htb]
    \centering
    \includegraphics*[width=75mm]{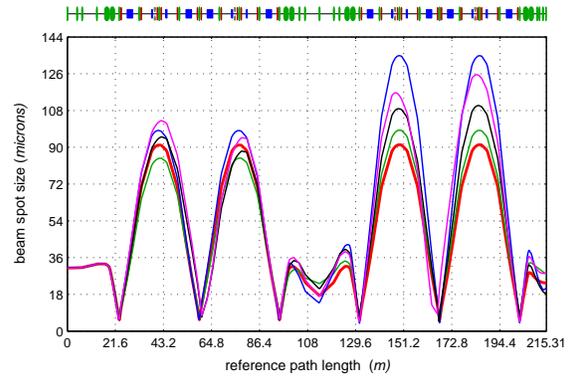}
    \caption{Evolution of beam spot size ($\sqrt{\sigma_x \sigma_y}$) 
    along the collimation section. 
    Beam energy 17.5 $GeV$. Normalized emittances 1.4 $mm \cdot mrad$.
    Red curve shows the design spot size (linear theory). 
    All other curves are results extracted from the tracking 
    simulations. A matched Gaussian beam at the entrance with
    $-3\%$ (blue) and $+3\%$ (green) energy offsets, with
    $40 \sigma_y$ transverse offset (black), and with both
    $-3\%$ energy and $40 \sigma_y$ offsets (magenta).
    Sextupoles are switched on.}
    \label{figSpotS}
\end{figure}

\begin{figure}[htb]
    \centering
    \includegraphics*[width=75mm]{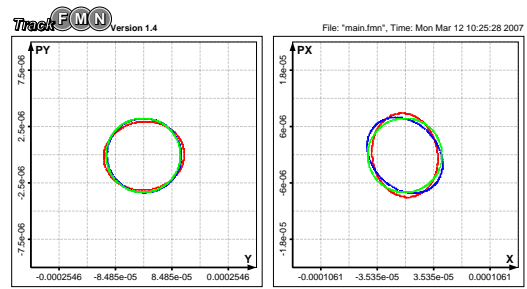}
    \caption{Phase space portraits of monochromatic
    3$\sigma_{x,y}$ ellipses (matched at the entrance) after 
    tracking through the entire collimation section. 
    The relative energy deviations are equal to $-1.5\%$, $0\%$ and $+1.5\%$ 
    (red, green and blue ellipses, respectively). 
    Sextupoles are switched on.}
    \label{figEll}
\end{figure}

The optics solution for the XFEL post-linac collimation section described 
in this paper meets all design specifications. It is capable of providing 
simultaneously a large beam spot size at the collimator locations 
(Fig.~\ref{figSpotS}) and, in the same time, to transport bunches with 
different energies (up to $\pm 1.5\%$ from nominal energy) while preserving
with good accuracy energy independent input and output matching conditions 
(Fig.~\ref{figEll}). These criteria are met by designing a magnetic system 
whose second-order chromatic and geometric aberrations are controlled by 
the symmetry of the first-order optics and sextupole fields.

The system uses four main primary collimators and the studies presented 
in~\cite{ColXFEL} show that these collimators are able to confine all particles 
which passed the collimation section freely (without touching collimator apertures) 
into a volume  in the phase space, that can be safely transported through all 
downstream beamlines (including undulator modules) to the beam dumps.

\end{document}